# Technical Notes:

# Shock-Induced Drop Size and Distributions

J.E. Park[1], T.-W. Lee[1*],

[1]Mechanical and Aerospace Engineering, SEMTE, Arizona State University, Tempe, Arizona, USA

**Abstract**

We use an integral analysis of conservation equations of mass and energy, to determine the drop size and distributions during shock-induced drop break-up. The result is an updated form for the drop size as a function of its final velocity, from a series of work applied to various atomization geometries. Comparisons with experimental data demonstrate the validity and utility of this method. The shock-induced drop size and distributions can be predicted within reasonable accuracy as a function of the drop velocity ratio and fluid properties. The result also illustrates the dynamical process of kinetic energy deficit transferred to the surface tension energy, and the skewing of the drop size distribution due to the non-linear dependence on velocity ratio.

*Corresponding author: T.-W. Lee, attwl@asu.edu

**Nomenclature**

A = cross-sectional area of the control volume containing the primary drop

C = constant relating the drop diameter to the atomization parameters (Eq. 4)

D = final drop diameter

$D_o$ = initial drop diameter

K, K′ = proportionality constants for the viscous dissipation term

n = drop number density

$u_b$ = final drop velocity

$u_o$ = initial drop velocity

V = volume of the spray bounded by A and spray length

$We_{\Delta u}$ = Weber number based on kinetic energy defect (Eq. 5)

**Greek Letters**

$\mu_L$ = liquid viscosity

$\rho_g$ = ambient gas density

$\rho_L$ = liquid density

σ = surface tension

## Introduction

Drop disintegration through shock interaction occurs in high-speed flows and propulsion devices. When a shock passes through a suspended droplet, the pressure differential and velocity change cause rapid distortion of the droplet surface, leading to various modes of fragmentation [1]. Examples include raindrop impingement on hypersonic aircrafts, supersonic combustion, and rotating detonation engines [2, 3]. The shock wave disintegrates the fuel drop into smaller ones, the evaporation rate of which tends to follow that of a representative drop diameter (Sauter mean diameter) [4]. Recent studies address this phenomenon through computational means [2-4]. However, due to small scales involved, full numerical resolution required to simulate atomization processes is excessive for typical droplets of 5 μm or below that are produced. The situation is also challenging for imaging experiments since the dynamic range from 5 μm to 1 mm droplet size necessitates high pixel density at fast framing rates. There are experimental data for the shock-induced droplet breakup. Recent work is by Chen et al. [5], in which water and liquid metal breakup by shock wave is considered. Droplet velocity, size and distributions are reported at various stages of the break-up. An extensive set of data goes back to 1990's, by Hsiang and Faeth [6, 7] who used the shock wave setup to investigate the secondary break-up. A correlation between the resulting drop size and velocity is reported in that work [6, 7].

In view of the recent interests on determination of the drop size and distribution, we have re-examined the integral analysis of secondary atomization from our prior work [8]. As described in a series of articles [8-11], the final drop size and distributions can be deduced with good accuracy by considering the conservation equations for a control volume enveloping the break-up process. This integral approach de-necessitates the heavy numerics by bypassing the complex, minute (differential) details of individual surface distortions leading to the final droplet state. Instead, as noted above since the overall mass evaporation rate of the droplet cloud is representable by a single mean-diameter droplet, we focus on finding this diameter as a function of the velocity ratios and fluid/gas properties. We present results on determination of drop size and distributions, based on updated integral analysis.

**Theoretical Formulation**

The drop size arising from various injection geometries has been considered by Lee and co-workers [8-11]. Secondary atomization, or drop fragmenting into smaller ones, has also been considered [8], leading to the following energy balance.

$$\rho_L \frac{u_o^3}{2} A = \frac{\pi}{12} n \rho_L u_b^3 D^3 + n u_b A \pi \sigma D^2 + K \mu_L \left(\frac{u_o - u_b}{D}\right)^2 V \qquad (1)$$

The initial kinetic energy (left-hand side) entering the control volume at $u_o$ is re-distributed to the final kinetic energy (first term on the right-hand side), surface tension energy, and viscous dissipation. Relative to our prior work [8], we make some modifications and improvements. The viscous dissipation is now heuristically written as the squared ratio of the velocity difference ($u_o$-$u_b$) and the initial drop size, $D_o$, which is equivalent to taking the velocity and length scales, respectively. Here, A is the cross-sectional area, and V the control volume enveloping the break-up process [8]. We take A and V to be proportional to square ($D_o^2$) and cube ($D_o^3$) of the parent (initial) drop diameter, respectively. K is a proportionality constant for the viscous dissipation. For the drop number density (n), we need a reference volume since n has the unit of inverse volume, which we again take to be proportional $D_o^3$. We also consider a representative mean drop size, D, during the secondary break-up. Then, Eq. 1 yields an expression for the ratio of final to initial drop diameter in Eq. 2.

$$\frac{D}{D_o} = \frac{6 u_b \sigma}{\frac{\rho_L D_o u_o^3}{2}\left(1 - \frac{u_b^3}{u_o^3}\right) - K' \mu_L u_o^2 \left(1 - \frac{u_b^2}{u_o^2}\right)} \qquad (2)$$

Eq. 2 is in the form of the final drop size relative to the initial, $D/D_o$ as a function of the velocity ratio, $u_b/u_o$. $u_b$ is the final drop velocity, relative to the initial, $u_o$. We can see that Eq. 2 provides the drop size in an analytical form, which results from the initial kinetic energy being re-distributed into the final kinetic, surface tension energy and viscous dissipation. We can also look at the experimental correlation (Eq. 3) by Hsiang and Faeth [7], which is obtained by fitting with the data. The original form of the correlation is:

$u_o/u_b = f(D/D_o)$, and Eq. 3 is an inverted version to obtain the drop size D as a function of $u_b/u_o$. In Eq. 3, the surface tension (Weber number) effect is not included.

$$\frac{D}{D_o} = \frac{(2.7)^{3/2}\left(\frac{\rho_g}{\rho_L}\right)^{1/2}}{\left(\frac{u_o}{u_b}-1\right)^{3/2}} \quad (3)$$

Due to high kinetic energy involved during shock interaction, the viscous dissipation effect tends to be quite small (negligible) [4, 6, 7]. Parametric changes in the viscosity hardly alter the drop size result in Eq. 2 at typical shock interaction conditions (M>1). This renders a generalized form for the drop size in Eq. 4. This expression can be written in a yet simpler expression using a Weber number, $We_{\Delta u}$, based on the "kinetic energy defect" (Eq. 5). C absorbs the numerical constants, and based on comparisons with experimental data, C=9, is recommended in Eq. 4.

$$\frac{D}{D_o} = C\frac{u_b}{u_o}\frac{\sigma}{\rho_L D_o u_o^2\left(1-\frac{u_b^3}{u_o^3}\right)} = C\frac{u_b}{u_o}\frac{1}{We_{\Delta u}} \quad (4)$$

$$We_{\Delta u} = \frac{\rho_L u_o^2\left(1-\frac{u_b^3}{u_o^3}\right)D_o}{\sigma} \quad (5)$$

Density ratio ($\rho_g/\rho_L$) is implicit through the velocity ratio in Eq. 4. Both the density and the final velocity ratio across are fixed across a shock wave for a given Mach number. However, $u_b/u_o$ in Eqs. 3 and 4 is the instantaneous, local velocity ratio. The momentum equation with estimated drag coefficient has been used to track the droplet velocity ratio by Hsiang and Faeth [7]. The gas and liquid densities appear in the drag and mass inertia terms, respectively. The velocity ratio, or the kinetic energy defect, is the key parameter in Eq. 4 since it is this reduction in kinetic energy that is transferred to the surface tension energy. This process is identical to other atomization processes where the aerodynamic interaction leads to the reduction in the kinetic energy to produce a large number of small droplets with high surface tension energy [8-11]. In this way, the total energy, momentum and mass are conserved in the spray or droplet flow field.

**Comparisons with Experimental Data**

The above analytical results lead to predictions of $D_{32}$, as a function of velocity ratio and fluid properties (Eq. 4). A comparison is shown in Fig. 1, with experimental data from Hsiang and Faeth [7], along with their correlation (Eq. 3). The authors used five different fluids [7], but for clarity in the plot we choose three (water, 63% glycol solution, and ethanol) in Fig. 1. We can see that the surface tension ($\sigma$) in the three different fluids has a small but finite effect, with larger $\sigma$ resulting in a larger drop size. The surface tension is 0.071 N/m (water), 0.065 (glycol solution), and 0.024 (ethanol). This is intuitively expected, and Eq. 4 puts in an analytical form. Current theory traces the experimental data quite well except for low velocity ratios (<0.4). It can be seen that the drop size is approximately 0.1 to 0.4 of the initial, for velocity ratio ranging from 0.4 to 0.8. Outside of this range, the drop size change becomes progressively more significant at limiting velocity ratio of $u_b/u_o \rightarrow 1$. $u_b/u_o \rightarrow 1$ corresponds to zero loss in kinetic energy so that the drop size becomes unchanged. Therefore, the limiting drop size or cut-off is $D/D_o = 1$ for $u_b/u_o \rightarrow 1$. The other limit of $u_b/u_o \rightarrow 0$ would mean all of the kinetic energy is lost, resulting in very small drop size. Thus, Eq. 4 accounts for this transfer of kinetic to surface tension energy, resulting in predictions of the drop size, as illustrated in Fig. 1. As noted above, the density effects are through the momentum balance, which is represented by the velocity ratio in Eq. 4. In Fig. 1, the experimental correlation of Hsiang and Faeth [7] start to deviate from the data for $u_b/u_o$ <0.4 and $u_b/u_o$ >0.6, when plotted in semi-log graph (the original version was $u_b/u_o$ =f($D/D_o$) plotted in log-log graph [7]).

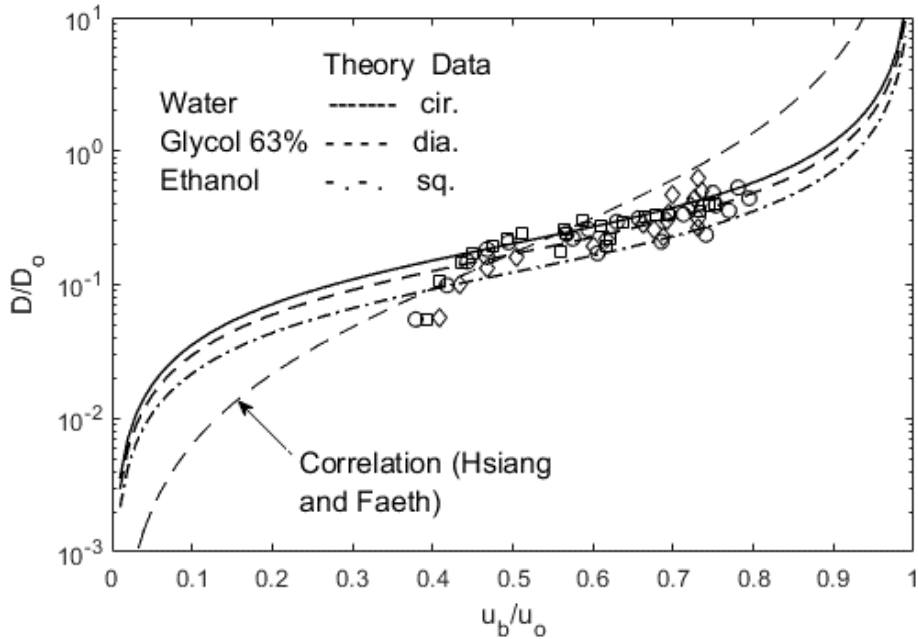

**Fig. 1. Comparison with current theory (Eq. 4) with experimental data/correlation by Hsiang and Faeth [7].**

Eq. 4 is in the form of drop size-velocity correlation, so that it can be used to transform the velocity to drop size distributions, as shown in Fig. 2. Comparisons are made experimental data of Chen et. al [5], who measured the drop size distributions for water and Galistan liquid droplets. We can see in Fig. 2 that the drop size distribution at the same Weber number (We=33) and drop location (x=25 mm) are quite similar between water and Galistan in spite of the large differences in the fluid properties. The liquid densities are 1000 and 6440 kg/m$^3$, respectively for water and Galistan, while the surface tension has nearly a factor of 10 difference (0.073 N/m for water; and 0.718 N/m for Galistan). These two fluid parameters act in opposite directions, surface tension increasing the drop size while the liquid density decreases it. However, the ratio of the surface tension increase is larger, so that it leads to a small, but observable shift toward larger drop size distribution in Fig. 2. The velocity distribution is not given in Chen et al. [5], so that we use an assumed clipped Gaussian probability density function with the observed mean velocity. Due to the non-linearity in the drop size as a function of the velocity, a symmetric velocity distribution is transformed to a skewed drop size distribution, resembling lognormal or Rosin-Rammler type [8-11].

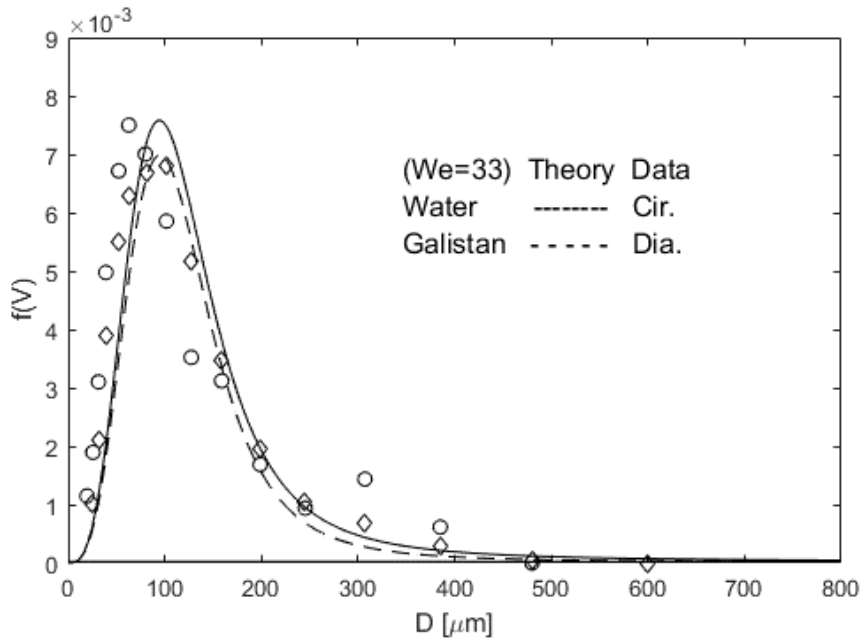

**Fig. 2. Comparison of the drop size distribution. Experimental data are from Chen et al. [5].**

Effects of various fluid properties can now be predicted using Eq. 4. As noted above, an increase in the surface tension would result in an increase the drop size, as confirmed in Fig. 3 (2σ line). Conversely, increase in the liquid density (1.5$\rho_L$ line) lowers the drop diameter, relative to the reference water drop size. The physical reason for this is that for the same velocity ratio, the kinetic energy defect is larger for higher liquid density. As noted above, the effect of the viscosity is small, requiring a 100-fold increase to exhibit an appreciable effect in Fig. 3. Nonetheless, all of the fluid and dynamic property effects can be quantitatively determined using Eq. 2 or 3. The experimental correlation by Hsiang and Faeth [7] does not take into account of the key parameter, surface tension. Figure 3 illustrates the reason that the experimental data [7] do not vary significantly with different fluids in Fig. 1. For example, the liquid density increases to 1126 kg/m³ for glycol mixture, but the surface tension decreases to 0.0648 N/m relative to water ($\rho_L$=998 kg/m³ and σ= 0.072 N/m). Since the liquid density and surface tension act in opposite directions and the property changes are relatively small, the resulting drop size is not significantly different. For ethanol, both the liquid density and surface tension decrease, relative to water. However, since the surface tension decrease is more significant, the resulting drop size is slightly smaller in Fig. 1. Current theory allows for determination of the drop size for any fluid, by inputting its fluid properties.

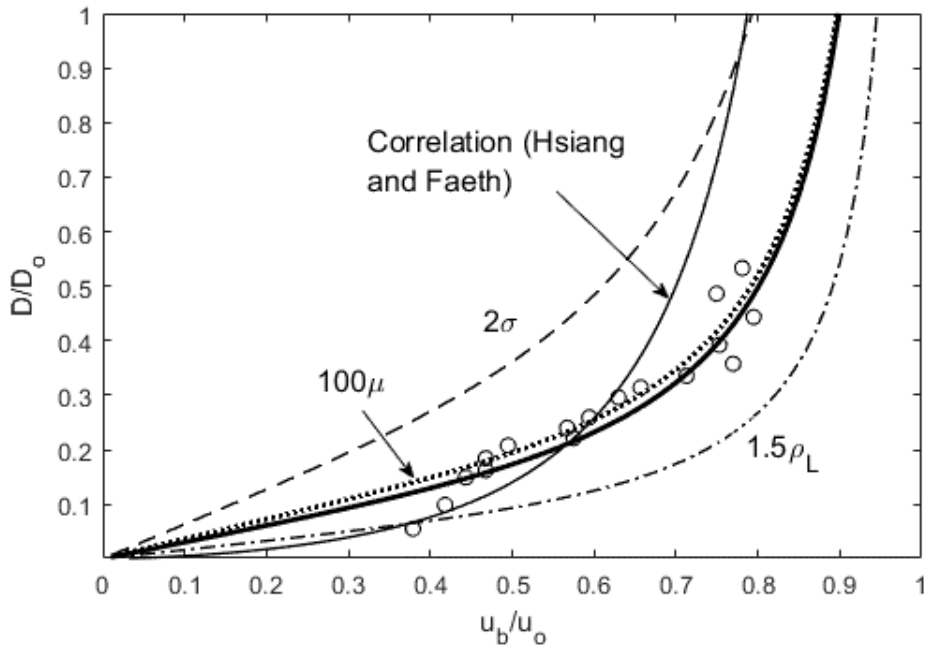

**Fig. 3. Effects of fluid property changes, relative to water. Experimental data and correlation by Hsiang and Faeth [7] are also plotted as a reference.**

**Conclusions**

Interests in shock-induced drop break-up have re-emerged due to applications in raindrop impingement on hypersonic aircrafts, supersonic combustion, and rotating detonation engines. Experimental and computational analyses are useful for providing key data; however, running them through a matrix test conditions and fluids can be prohibitively expensive and time-consuming. We have used some archival experimental data sets to validate an analytical method for determining the final drop size and distributions during shock-induced atomization of a primary (initial) drop. Integral analysis of the conservation of mass and energy leads to a closed-form equation for $D/D_o$ as a function the drop velocity ratio, $u_b/u_o$ (Eq. 4). This also serves as the drop size-velocity correlation, so that velocity distribution can be converted to drop size distributions.